\documentclass[12pt]{aipproc}
\usepackage{amsmath}

\newcommand{\braket}[1] {\langle #1\rangle}

\layoutstyle{6x9}

\begin{document}

\title{Results and Frontiers in Lattice Baryon Spectroscopy}

\author{John Bulava\footnote{Speaker, jbulava@andrew.cmu.edu}\thinspace}{address={Department of Physics, Carnegie Mellon University, Pittsburgh, PA 15213, USA}}
\author{Robert Edwards}{ address={Thomas Jefferson National Accelerator 
Facility, Newport News, VA 23606, USA}}
\author{George Fleming}{address={Yale University, New Haven, CT 06520, USA}}
\author{K. Jimmy Juge}{address={Department of Physics, University of the Pacific,
 Stockton, CA 95211, USA}}
\author{Adam C. Lichtl}{address={RBRC, Brookhaven National Laboratory, Upton, NY 11973, 
USA}}
\author{Nilmani Mathur}{address={Tata Institute of Fundamental Research, 
 Mumbai 40005, India}} 
\author{Colin Morningstar}{address={Department of 
Physics, Carnegie Mellon University, 
Pittsburgh, PA 15213, USA}}
\author{David Richards}{address={Thomas Jefferson National Accelerator 
Facility, Newport News, VA 23606, USA}}
\author{Stephen J. Wallace}{address={University of Maryland, 
College Park, MD 20742, USA}}

\begin{abstract}
The Lattice Hadron Physics 
Collaboration (LHPC) baryon spectroscopy effort is reviewed. 
To date the LHPC has performed exploratory Lattice QCD calculations of the 
low-lying spectrum of Nucleon and Delta baryons. 
These calculations demonstrate the effectiveness of our method by obtaining 
the masses of an unprecedented number of excited states with 
definite quantum numbers. Future work of the project is outlined.
\end{abstract}
\copyrightholder{}
\copyrightyear{}
\maketitle
A main goal of the LHPC is to determine the spectrum of excited baryons. 
This is achieved by extracting hadron masses from the exponential 
fall-off of two-point temporal correlation functions  
$\braket{0|\mathcal{O}_i(t)\bar{\mathcal{O}}_j(0)|0}$, in which the 
operators $\mathcal{O}_i$ are composed 
of quark and gluon fields. These correlators can be expressed in terms of path 
integrals on a discrete space-time lattice~\cite{BOOKS} which are 
calculated numerically using Monte Carlo techniques~\cite{ColinQFT}. 

The method developed by the 
LHPC to extract baryonic spectra is detailed in 
Refs.~\cite{ColinGT,Smring,AdamsThesis}.
This involves the construction of interpolating operators
to create states with 
definite quantum numbers, the evaluation of the temporal correlators 
$\braket{0|\mathcal{O}_i(t)\bar{\mathcal{O}}_j(0)|0}$,
and the pruning of the final operator sets. 

If we have a set of $N$ operators $\{\mathcal{O}_i\}$, we may form the 
$N\times N$ (time-dependent) matrix
\begin{equation}\label{corr}
C_{ij}(t) = \braket{0|\mathcal{O}_i(t)\bar{\mathcal{O}}_j(0)|0}.
\end{equation}
The diagonalization of $C^{-{1}/{2}}(t_0)C(t)C^{-{1}/{2}}(t_0)$, where 
$t_0$ is a small non-zero reference time, 
produces correlators that 
may be fit to decaying exponentials to obtain $E_i$, the energies of 
the lowest-lying states that can be 
interpolated by $\{\mathcal{O}_i\}$. If only zero momentum 
operators are used, the extracted energies of single particle states  
correspond to their masses. 

\begin{figure}[t]
\centerline{
\raisebox{3mm}{\setlength{\unitlength}{1mm}
\thicklines
\begin{picture}(16,10)
\small
\put(8,6.5){\circle{6}}
\put(7,6){\circle*{2}}
\put(9,6){\circle*{2}}
\put(8,8){\circle*{2}}
\put(4,0){single-}
\put(5,-3){site}
\end{picture}}
\raisebox{3mm}{\setlength{\unitlength}{1mm}
\thicklines
\begin{picture}(16,10)
\small
\put(7,6.2){\circle{5}}
\put(7,5){\circle*{2}}
\put(7,7.3){\circle*{2}}
\put(14,6){\circle*{2}}
\put(9.5,6){\line(1,0){4}}
\put(4,0){singly-}
\put(2,-3){displaced}
\end{picture}}
\raisebox{3mm}{\setlength{\unitlength}{1mm}
\thicklines
\begin{picture}(20,8)
\small
\put(12,5){\circle{3}}
\put(12,5){\circle*{2}}
\put(6,5){\circle*{2}}
\put(18,5){\circle*{2}}
\put(6,5){\line(1,0){4.2}}
\put(18,5){\line(-1,0){4.2}}
\put(6,0){doubly-}
\put(4,-3){displaced-I}
\end{picture}}
\raisebox{3mm}{\setlength{\unitlength}{1mm}
\thicklines
\begin{picture}(20,13)
\small
\put(8,5){\circle{3}}
\put(8,5){\circle*{2}}
\put(8,11){\circle*{2}}
\put(14,5){\circle*{2}}
\put(14,5){\line(-1,0){4.2}}
\put(8,11){\line(0,-1){4.2}}
\put(4,0){doubly-}
\put(1,-3){displaced-L}
\end{picture}}
\raisebox{3mm}{\setlength{\unitlength}{1mm}
\thicklines
\begin{picture}(20,12)
\small
\put(10,10){\circle{2}}
\put(4,10){\circle*{2}}
\put(16,10){\circle*{2}}
\put(10,4){\circle*{2}}
\put(4,10){\line(1,0){5}}
\put(16,10){\line(-1,0){5}}
\put(10,4){\line(0,1){5}}
\put(4,0){triply-}
\put(1,-3){displaced-T}
\end{picture}}
\raisebox{3mm}{\setlength{\unitlength}{1mm}
\thicklines
\begin{picture}(20,12)
\small
\put(10,10){\circle{2}}
\put(6,6){\circle*{2}}
\put(16,10){\circle*{2}}
\put(10,4){\circle*{2}}
\put(6,6){\line(1,1){3.6}}
\put(16,10){\line(-1,0){5}}
\put(10,4){\line(0,1){5}}
\put(4,0){triply-}
\put(2,-3){displaced-O}
\end{picture}}  }
\caption{The various types of extended baryon operators. Solid circles 
represent smeared quark fields, lines represent smeared 
link fields, and hollow circles the location of the reference site. 
\label{fig:operators}}
\end{figure}
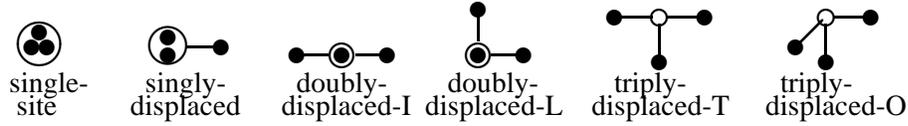

Our baryon operators are composed of covariantly displaced quark fields 
as building blocks combined in a gauge invariant way to 
interpolate both ground and excited baryonic states. 
Both the quark and link fields are \emph{smeared} (replaced
with a spatially localized weighted average)~\cite{Smring} to increase  
overlap with low-lying modes. 
Furthermore, they are designed to be 
\emph{spatially extended} in order to interpolate 
states which are radially and orbitally 
excited. This spatial extension is achieved by gauge-covariantly displacing 
some (or all) of the three quarks in different directions 
from some reference site. 
The various types of extended baryon operators are shown in 
Fig.~\ref{fig:operators}.

Group theoretical projections are used to construct operators which transform 
irreducibly under the symmetries of the lattice.
Recall that a baryonic state 
is denoted by its mass ($m$), spin and parity ($J^P$), and isospin ($I$). To 
construct states with definite isospin, we combine quark fields of different
flavors to create an operator that transforms irreducibly under SU(2) isospin. 
Since the QCD Lagrangian is diagonal in 
flavor (and thus isospin) space, members of an iso-multiplet will be mass 
degenerate. Thus we may arbitrarily choose the operator with maximal third 
component of isospin $I_3$ as the interpolating operator for a given 
iso-multiplet. 

In continuous
space, $J^P$ labels an irreducible representation of the 
(double-valued) rotation group SU(2), which is a symmetry group of  
rotationally invariant Hamiltonians. However, on a cubic lattice the symmetry 
group is no longer that of continuous rotations but the subgroup of discrete
lattice rotations. 
This group is denoted $O^D_h$ and is termed the double cubic point group. 
Because this group 
has a finite number of elements, it also has a finite number of irreducible
representations, in contrast to the infinite number of possible $J^P$ 
values in continuous space. These representations are labeled 
$\mathrm{G_{1g}, H_g, G_{2g}, G_{1u}, H_u,}$ and $\mathrm{G_{2u}}$ for historical reasons. There also exists 
a procedure known as \emph{subduction} to identify the states appearing in 
these irreducible representations with continuum $J^P$ values. 
For a complete review of the operator construction procedure, see 
Ref.~\cite{ColinGT}. 

After employing these considerations, we have operators which interpolate 
states with definite isospin, lattice spin, and parity. However, for a particular 
isospin (e.g. the Nucleons or Delta baryons) and lattice spin-parity value (such 
as $\mathrm{G_{1g}}$ or $\mathrm{H_g}$), 
this operator set is 
unmanageably large. It is necessary to optimize or `prune' 
this set down to a manageable number~\cite{AdamsThesis}. 

 The operator set is first pruned based on intrinsic noise. 
 This amounts to selecting the 
 ten least noisy operators within each operator type
 (single-site, singly-displaced, \emph{etc}.). These operators are determined
 by examining the diagonal correlators  
 $\braket{0|\mathcal{O}_i(t)\bar{\mathcal{O}}_i(0)|0}$. 
 We find that these least noisy operators  
 have good overlap with the low-lying states of interest and couple weakly 
 to the higher-lying energy modes of the theory. 

In addition, noise enters if the operator set 
produces a correlation matrix which is ill-conditioned. 
After the 
least noisy set of operators of each type is determined, the 
subset of those operators with the lowest condition number is chosen. 
This condition number is calculated for  
the normalized correlation matrix 
${C_{ij}(t')}/{\sqrt{C_{ii}(t')C_{jj}(t')}}$, where $C_{ij}(t)$ is 
defined in Eq. ~\ref{corr} and evaluated at some early time 
$t'\approx 3a_t$. In determining the optimal subset of operators, both  
condition number and operator type diversity are taken into consideration.

An exploratory calculation designed to test the effectiveness of the 
method has been performed for the Nucleon~\cite{AdamsThesis} and 
Delta baryons. 
To this end, calculations were carried out on a small 
volume ($\approx 1.2\mathrm{fm}$) lattice with a 
relatively coarse ($\approx 0.1\mathrm{fm}$)
lattice spacing. For computational simplicity, the pion mass for 
this calculation was unphysically large ($m_{\pi} \approx 700\mathrm{MeV}$) and the 
quenched approximation (omission of quark loops) was employed.
Also, these results were based on a low-statistics ensemble of 
200 configurations. 

\begin{figure}
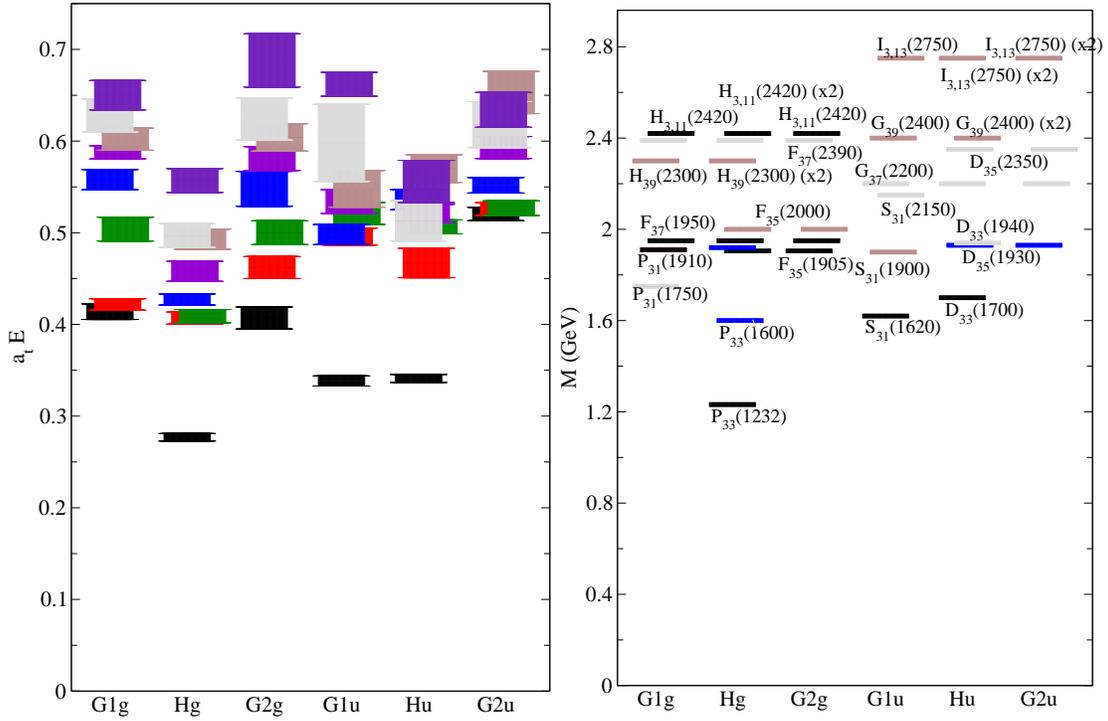
 
\begin{minipage}{.49\textwidth}
\includegraphics[width=\textwidth]{DeltaMasses.eps}
\end{minipage}
\begin{minipage}{.49\textwidth}
\includegraphics[width=\textwidth]{DeltaMasses_Expmnt.eps}
\end{minipage}
\caption{The low-lying Delta spectrum. The left plot is the lattice 
calculation (colors are used to distinguish different levels). In the experimental plot (right), black denotes a 4-star state, blue a 3-star
state, brown a 2-star state, and gray a 1-star state. 
States are grouped according to their lattice spin-parity values, 
$\mathrm{G_{1g}, H_g}$, etc.\label{fig:spectra}}
\end{figure}

In Fig.~\ref{fig:spectra}, the spectrum for each lattice $J^P$ value 
($\mathrm{G_{1g}, H_g}$, etc.) is plotted 
in lattice units. Comparison is made with experimental 
results~\cite{PDG} by identifying these lattice spin-parity values with their 
continuum 
counterparts using the group theoretical process of subduction. It should be 
noted that for $J\ge\frac{5}{2}$, one continuum state will have several lattice 
counterparts. 
The scales have been set using the mass of the lowest-lying hadron, the
$P_{11}(939)$. Shown here are the experimental and lattice results
for the Deltas. For an analogous plot of the Nucleons, see Ref. 
\cite{AdamCusco}. 

Upon examination of Fig. ~\ref{fig:spectra}, we notice that the lowest-lying energy state 
corresponds to a 
$J^P=\frac{3}{2}^+$ ($\mathrm{H_g}$) state, in agreement with the experimental 
$P_{33}(1232)$. The next two states occur in the 
$\frac{1}{2}^-$ ($\mathrm{G_{1u}}$)
and $\frac{3}{2}^-$ ($\mathrm{H_u}$) channels, in agreement with the experimental results. 
However, the first even parity excitation, the $P_{33}(1600)$, disagrees  
with the experimental result. This feature is also present in the nucleon 
sector~\cite{AdamCusco}, where the 
$P_{11}(1440)$ Roper resonance is higher than the first odd-parity excitation.
As our simulation neglected quark 
loops and possessed an unphysically high pion mass, this discrepancy indicates
that 
either these states are not interpolated by standard 3-quark operators (in the 
quenched approximation) or that they are sensitive to quark mass effects. 

The exploratory calculation described here will be followed by unquenched 
high-statistics runs at larger volumes, finer lattice spacings, 
and lower quark masses. Computational time has already been allotted for such 
runs using 10 million core-hours on the Cray XT3 at Oak Ridge 
National Lab and 3 million 
service units on the Cray XT3 `bigben' at the Pittsburgh Computing Center. 
  
Future work will utilize `all-to-all' 
propagators~\cite{All2all}, which estimate quark propagators from 
all initial sites 
to all final sites. This is in contrast with the currently used 
`point-to-all' 
propagators which 
only employ one initial site. This method will result in an increase in
statistics, but more importantly enable the construction of 
multi-hadron operators. 

In conclusion, this exploratory calculation of the low-lying Nucleon and Delta
spectra by the LHPC demonstrates the extraction of an unprecedented number of 
excited states in Lattice QCD. With the allocation of future computer resources
and the development of all-to-all propagators, our 
collaboration continues to move toward the ultimate goal of identifying states
with definite quantum numbers, 
predicting their masses, and calculating matrix elements.

This research is supported by NSF grant PHY 0653315 and
Dept. of Energy contracts DE-AC05-06OR23177 and 
DE-FG02-93-ER-40762. Travel to the VII Latin 
American Symposium on Nuclear Physics and Applications was made possible 
by the NSF Sponsorship Program.

\end{document}